\documentclass[11pt,a4paper]{article}

\usepackage{latexsym}
\usepackage{amssymb}
\usepackage{amsmath}
\usepackage[authoryear, round]{natbib}

\usepackage{ifpdf}

\ifpdf
    \usepackage[pdftex]{graphicx}
    \usepackage{color}
    \definecolor{myred}{rgb}{0.5,0,0}
    \definecolor{myblue}{rgb}{0,0,0.75}
    \definecolor{mygreen}{rgb}{0,0.5,0}
   \usepackage[pdftex,
               pdftitle={Capital allocation for credit portfolios under normal and stressed market conditions},
               pdfsubject={point-in-time PD, through-the-cycle PD, portfolio loss distribution},
               pdfkeywords={},
               pdfauthor={Norbert Jobst, Dirk Tasche},
               pdfstartview=FitR,
               breaklinks=true,
               colorlinks=true,
               citecolor=myred,
               linkcolor=myblue,
               urlcolor=mygreen]{hyperref}
\else
    \usepackage[dvips]{graphicx}
   \usepackage{hyperref}
\fi

\newtheorem{remark}{Remark}

\newtheorem{proposition}{Proposition}

\begin{document}
\title{Capital allocation for credit portfolios under normal and stressed market conditions}

\author{%
Norbert Jobst \& Dirk Tasche\footnote{%
Norbert Jobst works for Lloyds Bank, Dirk Tasche works for the UK Financial Services Authority.
The opinions expressed in this note are those of the authors and do not necessarily reflect views of their
respective employers.}}

\date{First version: September 27, 2010\\
This version: March 10, 2012}
\maketitle

\begin{abstract}
If the probability of default parameters (PDs) fed as input into a credit portfolio model are estimated as through-the-cycle (TTC) PDs stressed market conditions have little impact on the results of the capital calculations conducted with the model. At first glance, this is totally different if the PDs are estimated as point-in-time (PIT) PDs. However, it can be argued that the reflection of stressed market conditions in input PDs should correspond to the use of reduced correlation parameters or
even the removal of correlations in the model. Additionally, the confidence levels applied for the capital calculations might be made reflective of the changing market conditions. We investigate the interplay of PIT PDs, correlations, and confidence levels in a credit portfolio model in more detail and analyse possible designs of capital-levelling policies. Our findings may be of interest to banks that want to combine their approaches to capital measurement and allocation with active portfolio management which, by its nature, needs to be reflective of current market conditions.
\end{abstract}


\section{Introduction}\label{tas_sec_0}

Economic Capital (EC) is typically used within banks to
\begin{itemize}
	\item determine the amount of capital needed in order to ensure solvency,
	\item allocate actual capital to business units in a systematic manner, and
	\item facilitate origination activities by consistently assessing different opportunities within a RAROC
	(Risk Adjusted Return on Capital) framework.
\end{itemize}
Key ingredients to the computation of EC and RAROC are estimates for the underlying assets' probability of default (PD), loss-given-default (LGD), and correlation parameters.

Furthermore, when setting up an EC framework, it is commonly believed that these input parameters should be so-called through-the-cycle (TTC) estimates. In practice, the estimates are frequently derived as ``historic average''
estimates based on several years of observation resulting in fairly stable capital requirements for banks. 
Stable capital requirements are clearly desirable for a bank's longer term planning process and actual operations. The
Basel Committee on Banking Supervision confirmed this in their recent Basel~III recommendations 
where it identified pro-cyclicality of capital requirements as ``one of the most
destabilising elements of the crisis'' \citep{basel2010basel}. However, producing true TTC parameter estimates proves challenging.

In practice, banks often operate several PD and LGD models, and the extent to which these models display a true TTC behaviour varies considerably. In contrast to TTC estimates, point-in-time (PIT) estimates are based on current economic conditions and aim to provide more accurate, short term estimates of an obligor's default likelihood and losses. However, most models in practice are hybrid in nature, and, henceforth, the capital requirements as determined bottom up through usage of typical credit portfolio models (e.g. KMV, CreditMetrics) may vary accordingly and fail to reflect the true risk.

In this chapter, we investigate whether or not a stable level of economic capital can be achieved within a PIT modelling environment.
There is some literature on the properties of PIT and TTC PD estimates \citep[e.g.][]{Heitfield2005} but not much has been published on a definition
that can be used in practice to characterise PIT or TTC estimates. We therefore begin by considering and comparing
methods for PIT and TTC PD estimates and the connection between PIT PD estimates
and credit portfolio risk modelling.
It turns out that full consistency of PD estimation and portfolio modelling is
feasible in principle but unlikely to be achieved in practice. We then demonstrate the consequences of
being consistent and
failing to be consistent in portfolio risk modelling and PD estimation by a numerical example. 
We also demonstrate the effect of
one possible work-around, the ``time-varying target solvency probability'' as suggested by
\citet{Gordy&Howells06} in the context of stabilising regulatory capital requirements.


\section{TTC vs.\ PIT PD modelling}
\label{tas_sec_1}

TTC estimates are appropriate for stabilising capital requirements in the mid and long term.
Use of TTC estimates, however, risks to overlook immediate and near-term developments. Particularly in a crisis situation,
an institution might suffer significant losses when relying on TTC estimates only.
PIT estimates are appropriate when a realistic short to mid term view of risk is required. Ideally, PIT and TTC estimates would be consistent for longer horizons in that long-horizon PDs converge to TTC PDs whereas short-term estimates are more reflective of current economic conditions.

A risk measurement policy for a financial institution therefore could include the following elements:
\begin{itemize}
\item Capital requirements: Based on TTC estimates.
	\item Short term lending: Rely on PIT estimates and PIT risk assessments for decision making.
	\item Mid to long term lending: Rely on TTC estimates and TTC risk assessments for decision making although near-term fluctuations should be considered as indicated above.
	\item Credit extensions and limit management for existing obligors: Look at PIT estimates and PIT
	risk assessments to identify obligors who are temporarily high risk (e.g. with PIT PD of 5\% or more).
\end{itemize}

Key for the implementation of a policy with such elements is the ability to estimate PIT and TTC PDs.
\citet{Aguais&al2006} describe an approach to PIT PD modelling which is based
on a structural model and uses segment-wise average indices of the so-called distance-to-default
to infer the current state in the credit cycle. This approach is very attractive as it provides a way to
determine TTC PD estimates at the same time.
In addition, this approach ties in very naturally with the estimation
of a correlation structure for a credit portfolio model.

From a causality perspective -- why are obligors more vulnerable in certain time periods? --
PIT PD estimation approaches based on macroeconomic factors like GDP growth are also quite
interesting. \citet{Engelmann&Porath}, for instance, suggest that a proportional hazard rate model
can be utilised very efficiently to include macroeconomic factors and thus provide PIT PD
estimates.
	
The further discussion in this chapter draws mainly on the estimation approach suggested by
\citet{Hamerle&Liebig&Scheule}. This approach is based on logit or probit regression that includes
macreeconomic factors but also a latent factor to better capture correlation. We adopt here
the simpler version of the methodology without latent factor because it provides an intuitive
framework that allows a consistent approach to PIT and TTC PD estimation on the one hand and
credit portfolio risk modelling on the other hand.

The need for such a consistent approach is intensified by increased regulatory requirements for stress testing.
Clarity with respect to PIT and TTC
risk assessments is required to avoid ``double stressing'' and to more efficiently implement specific stress scenarios.

A policy to achieve consistency of PIT and TTC risk assessments could include the following elements:
\begin{itemize}
	\item Estimate PIT PDs with (e.g.) the \citet{Hamerle&Liebig&Scheule} probit approach.
	\item Transform PIT PDs into TTC PDs (e.g., as described in section~\ref{sec:modelling} below).
	\item Use the factor structure as provided by the probit regression as the
	factor model for the credit risk portfolio model. 
	Note that this is different to typical portfolio models where frequently factor models 
	based on equity or default correlations are used and estimated using long periods of historic data resulting in predominantly TTC correlation parameters.
	\item To study stress scenarios, evaluate certain factors or apply the distribution
	truncation approach as suggested by \citet{Bonti&Kalkbrener&Lotz&Stahl}.
\end{itemize}


\section{Portfolio credit risk modelling and PD estimation}
\label{sec:modelling}

The purpose of this section is to describe an ideal situation such that portfolio credit
risk modelling and PD estimation are perfectly consistent.
While such a situation is unlikely to be found in practice, knowing an example of how it could
look like nonetheless might help to improve consistency between the portfolio model in place and the way parameter
estimation is done.
In addition, the simple setting of this section should help to clarify the notions of
point-in-time (PIT) and through-the-cycle (TTC) PDs.


\subsection{A simple credit portfolio model}
\label{sec:portfolio}

For the purpose of illustration we consider
a simple default-mode Gaussian copula credit portfolio model with deterministic loss severities.
In this model, the portfolio-wide
loss realised in one observation period can be described as follows:
\begin{equation}\label{eq:model}
\begin{split}
	L & = \sum_{i=1}^N u_i\,\mathbf{I}(D_i) \\
	D_i & = \bigl\{\sqrt{1-\varrho_i^2}\,\xi_i + \varrho_i\,w_i'S \le T_i\bigr\}
\end{split}	
\end{equation}
Explanation of the notation used in \eqref{eq:model}:
\begin{itemize}
	\item $L$ is portfolio-wide realised loss if we assume that losses occur only as a
	consequence of default and loss given default is deterministic.
	\item $N$ is the number of obligors in portfolio.
	\item $u_i$ is the (deterministic) loss in case of default associated with obligor $i$. In this chapter, we
	consider relative losses, i.e.\ we assume $\sum u_i = 1$.
	\item $D_i$ denotes the event that obligor $i$ defaults.
	\item $\mathbf{I}(A)$ is the indicator function of the event $A$, i.e.\ $\mathbf{I}(A) = 1$ if $A$ occurs
	and $\mathbf{I}(A) = 0$ otherwise.
	\item $S = (S_1, \ldots, S_k)$ is a random vector of \emph{systematic factors}, assumed
	to be multi-variate normal with mean vector 0. The number $k$ of factor usually is
	small (e.g.\ not greater than 10).
	The factors could be GDP growth, change in unemployment rate, or something similar.
	\item $w_i = (w_{i 1}, \ldots, w_{i k})$ is the vector of weights of the systematic factors
	associated with obligor~$i$. The weights are
	assumed to be normalised such that $\mathrm{var}[w_i'S] =1$ ($w_i'S = \sum_{j=1}^k w_{i j}\,S_j$
	denotes the Euclidian inner product).
	\item $\xi_i$ is the \emph{individual risk factor} associated with obligor $i$, assumed to be
	standard normal and stochastically independent of all other random variables in the model.
	$X = (\xi_1, \ldots, \xi_N)$ denotes the vector of all $\xi_i$.
	\item $\varrho_i \in [-1, 1]$ denotes the \emph{sensitivity} of obligor $i$ to the systematic risk.
	$\varrho_i = 0$ means that the obligor $i$'s wealth is not vulnerable with regard to systematic
	risk (obligor is acyclical). $\varrho_i = 1$ means that obligors $i$'s wealth is completely
	determined by the systematic risk. $\varrho_i < 0$ would imply that obligor $i$ is counter-cyclical.
	\item $T_i$ is obligor $i$'s \emph{default threshold}. $T=(T_1, \ldots, T_N)$ denotes the vector
	of all $T_i$.
\end{itemize}
Usually only the individual factors $\xi_i$ and the vector $S$ of systematic factors are
considered random elements. The default threshold $T_i$ is determined individually for
each single obligor but often not considered random although it may change year on year if the
obligor's perceived creditworthiness changes. \emph{In this chapter we allow for $T_i$ to be
a random variable as this helps to consider some issues more rigorously.}

\paragraph{TTC portfolio risk measurement.}
The distribution of the portfolio loss $\mathrm{P}[L \le \ell]$, $0 < \ell < 1$, usually is
estimated from an artificially generated sample of loss realisations
-- by Monte-Carlo simulation of realisations of the
individual risk factors
$\xi_i$ and the systematic factors $S$ but usually not the thresholds $T_i$.
This process is equivalent to trying to calculate the
probabilities $\mathrm{P}[L \le \ell]$ by taking the expectation of the event $\{L \le \ell\}$
with respect to $X$ and $S$. Technically speaking, as the loss $L = L(X, S, T)$ is a function
of the random vectors $X$, $S$, and $T$, taking the expectation results in
\begin{equation}\label{eq:int}
	\int \mathbf{I}_{[0,\ell]}\bigl(L(x,s,T)\bigr)\,\mathrm{P}_{(X,S)}(d x, d s)\ =\
	\mathrm{P}\bigl[L(X, S, t) \le \ell\bigr]\bigm|_{t=T}.
\end{equation}
The right-hand side of \eqref{eq:int} is not the unconditional probability $\mathrm{P}[L \le \ell]$ but
-- if we assume that the risk factors $X$ and $S$ and the thresholds $T$ are stochastically
independent -- can be interpreted as
the conditional probability $\mathrm{P}[L \le \ell\,|\,T]$. The assumption
of stochastic independence is crucial for this interpretation. In general the formula
for the conditional probability $\mathrm{P}[L \le \ell\,|\,T]$ is given by
\begin{equation}\label{eq:int2}
		\mathrm{P}[L \le \ell\,|\,T]\ = \
		\int \mathbf{I}_{[0,\ell]}\bigl(L(x,s,T)\bigr)\,\mathrm{P}_{(X,S)\,|\,T}(d x, d s).
\end{equation}
This differs from the left-hand side of \eqref{eq:int} by the fact that the integration
is done with respect to the conditional distribution $\mathrm{P}_{(X,S)\,|\,T}$ of
$(X, S)$ given $T$ which in general -- unless $(X, S)$ and $T$ are stochastically
independent -- is not the same as the unconditional distribution $\mathrm{P}_{(X,S)}$
of $(X,S)$. But it is the unconditional distribution of $(X,S)$ that is approximated
by the Monte-Carlo simulation. \emph{Hence Monte-Carlo simulation of $(X,S)$ with fixed $T$
as described above yields an economically interpretable result\footnote{%
Note that for this observation to be valid no independence assumption for
the systematic factors $S$ and the idiosyncratic factors $X$ needs to be made. For practical
applications, nonetheless it is common to assume independence.} if and only if the risk factors
$(X,S)$ and the threshold $T$ can be assumed to be stochastically independent.}

Note that integrating with respect to the systematic factors
means that a cycle-neutral (or TTC) perspective on the risks is adopted. Clearly, TTC PDs
should be used for this kind of calculation.

The random variables in the definition of the default event $D_i$ in \eqref{eq:model} are normalised in
such a way that $\sqrt{1-\varrho_i^2}\,\xi_i + \varrho_i\,w_i'S$ is standard normal and hence
the TTC PD of obligor $i$ -- more correctly the PD of obligor $i$ conditional on $T_i$ which is
cycle-neutral after integration with respect to $S$ -- is\footnote{$\Phi$ denotes the standard normal
distribution function.}
\begin{equation}\label{eq:TTC_PD}
	PD_i = \mathrm{P}[D_i\,|\,T_i]=\Phi(T_i).
\end{equation}

\paragraph{PIT portfolio risk measurement.}
In the model framework given by \eqref{eq:model}, a PIT view on the actual
risk of the portfolio can be achieved in two ways:
\begin{itemize}
	\item By taking the expectation of the event $\{L \le \ell\}$, $0 < \ell < 1$,
with respect to $X$ and $S$ as described above for TTC risk measurement, but
with a restricted range for $S$ to specify specific economic conditions. Details
about this approach can be found in \cite{Bonti&Kalkbrener&Lotz&Stahl}.
	\item By fixing a realisation $s$ of the systematic risk factor vector $S$ and taking
	the expectation of the event $\{L \le \ell\}$, $0 < \ell < 1$,
with respect to $X$ only -- while keeping the values of the $T_i$ constant at the same time.
As the vector $X$ of individual risk factors is assumed to be independent of
$S$ and $T$, this procedure gives estimates of the conditional distribution
$\mathrm{P}\bigl[L \le \ell\,|\,S,T\bigr]$, $0 < \ell < 1$.
\end{itemize}
In section \ref{sec:probit} we will see that in both approaches for the PIT risk view TTC PDs
should be used as input to the portfolio model.


\subsection{The probit regression model for PD estimation}
\label{sec:probit}

We describe in this section an approach to PD modelling and estimation that is perfectly consistent
with the portfolio model \eqref{eq:model}.
The approach we have chosen is essentially a simplified version of the probit
estimation suggested by \citet[][without latent factor]{Hamerle&Liebig&Scheule}.

We consider obligors in a fixed segment, defined say by industry or region. With each obligor~$i$,
we associate a vector of \emph{characteristic factors} $F_i = (F_{i 1}, \ldots, F_{i m})$ ($m$ does
not depend on $i$), $i = 1, \ldots, N$. Typically, $F_{ij}$ will be a financial ratio from the
balance sheet or a qualitative score like management quality.
Denote by $F$ the random matrix $F = (F_1, \ldots, F_N)$.

The point-in-time PDs resulting of a probit regression of default event indicators on the risk factors $F_i$ (different
for different obligors) and the vector of systematic factors $S$ (the same vector for each
obligor) can then be characterised as
\begin{equation}\label{eq:probit}
	PD_i(F, S) \ = \ \Phi(a_0 + a'F_i + b'S),
\end{equation}
with $a_0$ being a constant and $a$ and $b$ being constant vectors of appropriate dimension
(estimated by regression).

To see the connection between the probit model \eqref{eq:probit} and the Gaussian copula
credit portfolio model \eqref{eq:model} it is convenient to rewrite \eqref{eq:probit} as
a probability conditional on the factor vectors $F_i$ and $S$. Denote by $\xi_i$ a standard
normal random variable as in subsection~\ref{sec:portfolio} (with the same interpretation
as individual risk factor). Then we have
\begin{equation}\label{eq:cond_PD}
\begin{split}
	PD_i(F, S) & = \mathrm{P}\bigl[\xi_i \le a_0 + a'F_i + b'S\,|\,F_i, S\bigr]\\
	& = \mathrm{P}\left[\frac{\xi_i -b'S}{\sqrt{1+\mathrm{var}[b'S]}} \le
			\frac{a_0 + a'F_i}{\sqrt{1+\mathrm{var}[b'S]}} \,\Big|\,F_i, S\right]\\
	& = \mathrm{P}\bigl[\sqrt{1-\varrho^2}\,\xi_i + \varrho\,w'S \le
	T_i\,|\,F_i, S\bigr],
\end{split}
\end{equation}
where
\begin{itemize}
	\item the sensitivity $\varrho$ is defined by $\varrho =
		\frac{\sqrt{\mathrm{var}[b'S]}}{\sqrt{1+\mathrm{var}[b'S]}}$;
	\item the weight vector $w$ is defined by $w = -\frac{b}{\sqrt{\mathrm{var}[b'S]}}$;
	\item the default threshold $T_i$ is defined by
\begin{equation} \label{eq:TTC_threshold}
	T_i = (a_0 + a'F_i)\,\sqrt{1-\varrho^2} =
		\frac{a_0 + a'F_i}{\sqrt{1+\mathrm{var}[b'S]}}.
\end{equation}
\end{itemize}
\begin{proposition}\label{pr:one2one}
Assume that the same vector of systematic factors $S$ is used
both for portfolio risk modelling and for PIT PD estimation and that the constants in the description of the
default event $D_i$ in (\ref{tas_sec_0}) are not allowed to differ between different obligors but only between
different segments of obligors.
Then there is a one-to-one relation between the constants $\varrho$ and $w$ and the thresholds $T_i$ in the
default events $D_i$ in \eqref{eq:model} on the one hand and $b$, $\mathrm{var}[b'S]$, and
$(a_0 + a'F_i)_{i=1,\ldots,N}$ on the other hand.
\end{proposition}
\textbf{Proof.} We have seen above how $\varrho$, $w$ and the thresholds $T_i$ are determined by $b$, $\mathrm{var}[b'S]$, and
$a_0 + a'F_i$. Conversely, if $\varrho$, $w$ and the thresholds $T_i$ are given then $b$, $\mathrm{var}[b'S]$, and
$a_0 + a'F_i$ can be determined by
\begin{equation}
	\begin{split}
		\mathrm{var}[b'S] & = \frac{\varrho^2}{1-\varrho^2}\\
		b & = - w\,\frac{\varrho}{\sqrt{1-\varrho^2}}\\
		a_0 + a'F_i & = \frac{T_i}{\sqrt{1-\varrho^2}}.
	\end{split}
\end{equation}
This provides the other direction of the one-to-one relation. \hfill $\Box$

While we have shown how to parameterise the portfolio model \eqref{eq:model} from probit estimators
\eqref{eq:probit} and to derive a PIT probit estimator from the parameterisation of \eqref{eq:model},
to show full consistency of the model
\eqref{eq:model} and the probit PD estimator \eqref{eq:probit} we have to prove that indeed
the threshold $T_i$ from \eqref{eq:TTC_threshold} is related to a TTC PD that
derives in an intuitive way from the PIT PD \eqref{eq:cond_PD}.
The following
proposition shows that this consistency property of the portfolio model and the probit PD estimation can indeed be obtained
if the risk factors $F_i$ and the systematic factors $S$ are independent (i.e.\ the risk factors
are through-the-cycle).

\begin{proposition} \label{pr:PIT2TTC}
Assume that the risk factors $F_i = (F_{i 1}, \ldots, F_{i m})$ and the systematic
factors $S = (S_1, \ldots, S_k)$ are stochastically independent. Then the TTC PD corresponding
to \eqref{eq:probit} by integration with respect to the systematic factor vector $S$
is given by \eqref{eq:TTC_PD} with threshold $T_i$ as
defined by \eqref{eq:TTC_threshold}.
\end{proposition}
\textbf{Proof.} The TTC PD corresponding to a PIT PD is derived from the PIT PD by integration
with respect to the systematic factors over the full range they can take on. By independence of
$F_i$ and $S$ and the fact that $b'S$ is normally distributed we can calculate the TTC PD for
\eqref{eq:probit} as follows
\begin{align*}
	PD_i^{\mathrm{TTC}}(F, S) & =  \int PD_i(F,x)\, P_S(d x)\\
	& = \int \Phi(a_0 + a'F_i + y)\, P_{b'S}(d y) \\
	& = \int \mathrm{P}\bigl[\xi_i \le a_0 + a'F_i + b'S\,|\,b'S=y\bigr]\,P_{b'S}(d y)\\
	& = \mathrm{P}\bigl[\xi_i - b'S\le a_0 + a'F_i]\\
	& = \Phi\left(\frac{a_0 + a'F_i}{\sqrt{1+\mathrm{var}[b'S]}}\right).	
\end{align*}
In this calculation the random variable $\xi_i$ is assumed to be standard normal
and independent of all other random variables.
By definition of the default threshold $T_i$, we then have
$\Phi\Big(\frac{a_0 + a'F_i}{\sqrt{1+\mathrm{var}[b'S]}}\Big) = \Phi(T_i)$. This completes the proof.
\hfill $\Box$

In practice, integration of \eqref{eq:probit} with respect to the vector $S$ of systematic factors might
be approximated by averaging the values of \eqref{eq:probit} across a time series with realisations
of the systematic factors in both benign and poor economic conditions.

At first glance, the result of proposition \ref{pr:PIT2TTC} seems more or less self-evident.
However, in practice most of the time not all assumptions underlying proposition \ref{pr:PIT2TTC}
are satisfied such that consistency of portfolio modelling and PIT PD estimation might not be obtained.
\begin{itemize}
	\item The vector of systematic factors $S$ might not be normally distributed in reality.
	\item PIT PDs are not necessarily estimated by probit estimation but by logit estimation
	or other methods \citep[see, e.g.][]{Tasche2009a}. While this could be accounted for in the portfolio
	model by choosing another distribution for the individual risks, nonetheless a lot of further
	adjustments would be necessary which would make the model much less straight-forward.
    \item A bank typically operates several different PD models, some of which may be expert driven and less suitable for statistical modeling.
	\item It is unlikely that the characteristic factors $F_i$ and the systematic factors $S$ are
	stochastically independent. As a consequence, the portfolio loss distribution $\mathrm{P}[L \le \ell\,|\,T]$
	when calculated according to \eqref{eq:int}
	might be upward or downward biased, depending on the position in the credit cycle when the calculation
	is conducted.
\end{itemize}

\begin{remark} There are two other interesting consequences of proposition \ref{pr:one2one} on the one hand and proposition \ref{pr:PIT2TTC} and \eqref{eq:TTC_threshold} on the other hand.
\begin{enumerate}
	\item The PIT PD estimates according to \eqref{eq:probit} should never be used as an
	input to a model like \eqref{eq:model}, even if the model is applied to make a PIT risk measurement unless all other model parameters can be specified consistently as outlined in footnote \ref{ft:PIT} below.
	\item The naive transformation of the PIT PD \eqref{eq:probit} into a TTC PD by just removing the systematic
factors $S$ from the right-hand side of the equation is inadequate. The simple term $\Phi(a_0 + a'F_i)$
would underestimate the true TTC PD which according to \eqref{eq:TTC_threshold} and proposition \ref{pr:PIT2TTC}
is given by $\Phi\bigl((a_0 + a'F_i)\,\sqrt{1-\varrho^2}\bigr)$ (it would be an underestimation as
$a_0 + a'F_i$ should be negative most of the time for even PIT PDs in general will be less than 50\%). 
\end{enumerate}
\end{remark}
In this section we have discussed how -- under certain assumptions -- a consistent framework of PIT PD estimators, TTC PD estimates, and the factor model 
for the correlation structure of a credit portfolio model can be constructed. In particular, by 
proposition~\ref{pr:one2one} we have shown that there is a one-to-one mapping between suitable probit PIT PD estimators and
the factor model for the correlation structure while proposition~\ref{pr:PIT2TTC} provides a recipe of how
to derive TTC PD estimates from the PIT PD estimators such that the results can be used as input to the portfolio model.

In the following section, we demonstrate the potential issues with mixing up PIT and TTC parameters in standard portfolio models by means of a stylized example in the context of the computation of tail risk and EC. More 
specifically, the setting we consider is given by a simple one-factor special case of 
equations \eqref{eq:model} and \eqref{eq:TTC_PD}. 
In this setting, the mechanics provided by proposition~\ref{pr:one2one} is used to calculate the PIT PD values
and the conditions for proposition~\ref{pr:PIT2TTC} are satisfied because we work with an empty set of 
characteristic factors.


\section{Numerical example}
\label{sec:num}

We look at a stylised homogeneous portfolio with 100 assets. All assets have the same exposure, all LGDs
are assumed to be 100\%. We study the two cases of all assets being of investment grade quality (TTC PD 0.3\%)
and all assets being of sub-investment grade quality (TTC PD 3\%). The calculations
are done with a one-factor version of model \eqref{eq:model} and a sensitivity of 50\%
unless explicitly stated otherwise. As the calculation is for illustrative
purposes only we need not identify the systematic factor. We calculate\footnote{%
Thanks to the homogeneous one-factor setting the calculations can be done
by means of numerical integration -- no Monte-Carlo simulation is needed.} tail risk as Value-at-Risk (VaR) 
and economic capital (EC) as \emph{Unexpected Loss} which is
defined as \emph{VaR minus Expected Loss}.

The upper panel of table~\ref{tab:1} shows the results for the TTC case where the input PDs for the
calculation are just the TTC PDs assigned to the assets. VaR and EC here are calculated
at 99.9\% confidence level. Note that the confidence
level of 99.9\% implies a target TTC PD of 0.1\% for the bank holding the portfolio.
The VaR and capital values resulting from
the calculations seem rather high -- this is a consequence of the high granularity of the
portfolio, the 100\% LGD assumption and the relatively high sensitivity of 50\% to the
systematic factor.

The lower panels of table~\ref{tab:1} show the results for the PIT case where the input PDs
for the calculation are a) PIT PDs determined by having the single factor taking on the value -2.33
(which corresponds roughly to a 1 in 100 scenario) and b) the TTC PDs as in the TTC analysis.
For a), the TTC PD of 0.3\% is transformed
into a PIT PD of 3.4\% whereas the TTC PD of 3\% is transformed into a PIT PD of 20.4\%.
Otherwise the calculation is done exactly in the same way as for the TTC analysis -- that is why a) is
labelled ``PIT input, TTC calculation''. For b), TTC PDs are used as input PDs\footnote{%
Actually, it might be easier to implement this calculation by using the PIT PD derived as described
in a) as input PD and then run the portfolio model in TTC mode but with 
sensitivity $\varrho=0$.\label{ft:PIT}} to the model but
the calculation is done by integration with respect to the individual risk factors only while
the systematic risk factor has the fixed value -2.33. This is labelled ``TTC input, PIT calculation''
in table~\ref{tab:1}.
\refstepcounter{table}
    \begin{center}
\label{tab:1}
\parbox{15cm}{Table \thetable: \emph{Value-at-Risk (VaR) and economic capital (EC) 
results for a homogeneous portfolio
with 100 assets calculated with a one-factor model under TTC and PIT assumptions. EC is
determined as 'Unexpected Loss = VaR minus Expected Loss'. Details on input parameters
and calculation methods are provided in section~\ref{sec:num}.}}\\[2ex] 
\begin{tabular}{|l||c|c|}
\hline\hline
\multicolumn{3}{|c|}{\textbf{TTC analysis}}   \\ \hline\hline
Asset PD (TTC) & 3\%   & 0.3\%   \\ \hline
VaR (99.9\%) & 37\%   & 9\%  \\ \hline
Capital (99.9\%) & 34\%   & 8.7\%  \\ \hline\hline
\multicolumn{3}{|c|}{\textbf{PIT analysis: PIT input, TTC calculation}}   \\ \hline\hline
Input PD &	20.4\%	&	3.4\%	 \\ \hline
VaR (99.9\%) & 81\%   & 39\%  \\ \hline
TTC capital (99.9\%)	&	60.6\%	&	35.6\%  \\ \hline
VaR (98.7\%) & 64\%   & 21\%  \\ \hline
PIT capital (98.7\%)	 &	43.6\% 	&	17.6\%  \\ \hline
VaR (98\%) & 60\%   & 18\%  \\ \hline
PIT capital (98\%)	 &	39.6\% 	&	14.6\%  \\ \hline\hline
\multicolumn{3}{|c|}{\textbf{PIT analysis: TTC input, PIT calculation}}   \\ \hline\hline
Input PD &	3\%	&	 0.3\% \\ \hline
VaR (99.9\%) & 34\%   & 10\%  \\ \hline
TTC capital (99.9\%) &	13.6\%	&		6.6\% \\ \hline
VaR (98.7\%) & 30\%   & 8\%  \\ \hline
PIT capital (98.7\%)	 &	9.6\%	&		4.6\% \\ \hline
VaR (98\%) & 29\%   & 7\%  \\ \hline
PIT capital (98\%)	 &	8.6\%	&	3.6\% \\ \hline
\hline
\end{tabular}
\end{center}

The rows labelled ``VaR (99.9\%)'' and ``TTC capital (99.9\%)'' respectively 
in the lower panels of table~\ref{tab:1} display
the VaR and capital values calculated for these input PDs if the confidence level of 99.9\% remains
the same as for the TTC analysis. With PIT input PDs the new VaR and capital values are then clearly much
higher (e.g., for EC we have 60.6\% to 34\% for sub-investment grade and 35.6\% to 8.7\% for investment grade)
than the ones calculated for the TTC analysis. This picture changes dramatically when
the input PDs are the same TTC PDs as before in the TTC analysis (e.g., for EC 13.6\% to 34\% for
sub-investment grade and 6.6\% to 8.7\% for investment grade). Note, however, that
the VaR results are remarkably similar to the VaR results from the TTC analysis (34\% to 37\%
for sub-investment grade and 10\% to 9\% for investment grade). The reason for these significant
changes is the fact that in the PIT analysis default events become stochastically independent
because the calculations are conducted with a fixed value (-2.33) for the systematic factor.
Despite the higher input PDs the model \eqref{eq:model} therefore generates lighter tail
loss distributions than in the TTC analysis. In addition, expected loss is higher because
it is based on the PIT PDs.

The rows labelled ``VaR (98.7\%)'' and ``PIT capital (98.7\%)'', and 
``VaR (98\%)'' and ``PIT capital (98\%)'' respectively
in the lower panels of table~\ref{tab:1} display
the VaR and capital values calculated for the  PIT analysis if the confidence level is adapted to
reflect '1 in 100' PIT target PDs for the bank holding the portfolio. The target PD of
$1.3\% = 100\% - 98.7\%$ has been derived from the bank's TTC target PD of 0.1\%
in the same way as the PIT transformed PDs of the assets
in the portfolio. In particular the same sensitivity of 50\% to the systematic factor has been applied.
The target PD of
$2\% = 100\% - 98\%$ has been derived from the bank's TTC target PD of 0.1\%
under the assumption of a higher sensitivity of 71\% ($= \sqrt{50\%}$) to the systematic factor in order to reflect
the fact that the systematic risk of banks is higher than the systematic risk of most other industries.

Comparison of the most upper and the most lower panels of table~\ref{tab:1} shows that in the PIT analysis
the reduced uncertainty by the realisation of the systematic factor implies
that the portfolio model effectively is evaluated with stochastically independent
default events and, therefore, indicates broadly the same VaR values but less capital requirements than in the TTC analysis
even in a 1 in 100 stressed environment.
Adopting a PIT target PD for the bank holding the portfolio further enlarges
the difference between PIT capital figures and TTC capital figures, in particular
if the higher systematic risk of the bank is taken into account (3.6\% and 8.7\% for
investment grade and 8.6\% and 34\% for sub-investment grade). In contrast, PIT VaR figures
differ not too much from the TTC analysis VaR figures even if PIT target PDs are used to
determine the confidence level for VaR.

In total, the results of table \ref{tab:1} appear somewhat unintuitive as most people would
expect to see higher capital requirements in a 1 in 100 stressed environment. However, the relatively
low capital requirements are primarily a consequence of the fact that in the stressed
environment the expected loss that is deducted from the portfolio VaR to determine the unexpected loss
is very high because it is driven by the PIT PDs. 
Nonetheless, one might think again about whether a ``conditional independence'' (conditional on the
systematic factors) model like \eqref{eq:model} is really appropriate for the calculation
of risk in poor economic conditions. Weakening the conditional independence by
taking recourse to a latent systematic factor as suggested by \citet{Hamerle&Liebig&Scheule}
could be a work-around. In addition or alternatively, inclusion of contagion effects in the model structure
appears as a promising approach to improved modelling with regard to the issue of
unintuitive results.

The observations from table \ref{tab:1}  also confirm that
``time-varying target solvency probabilities'' \citep{Gordy&Howells06} indeed do
not only reduce regulatory capital requirements but also economic capital -- at least when they
are combined with a consistent approach to PIT portfolio risk modelling.


\section{Conclusions}

We have looked in some detail at an ideal framework of a credit portfolio model and a methodology for
the estimation of PIT PDs. In this framework, the same systematic factors are included in the portfolio
model and in the regression approach that is applied for PD estimation. There is actually a one-to-one
relation between the key portfolio model parameters and factors and the key PIT PD estimation parameters and factors.
It turns out, however, that stochastic independence of the
obligor-characteristic factors in the PD estimates and the systematic factors is a crucial condition both
for the portfolio model to be able to generate unbiased TTC risk measures and the portfolio model and
the PIT PD estimates being consistent in the sense that the TTC PD input parameters for the portfolio model
can be derived from the PIT PD estimates.

We have then demonstrated the conclusions from the theoretical considerations by a numerical example.
The example clearly shows that using PIT PD estimates as input to a portfolio model that is run in TTC mode
leads to unrealistically high capital requirements. The example shows also that this effect can be mitigated
to some extent by replacing the TTC confidence level for the capital calculation by a PIT confidence level
that is derived from the bank's own PIT PD.

In addition, the results of the numerical example indicate that PIT capital levels may come out unintuitively
low when calculated in the theoretically correct way. This is primarily a consequence of the fact that 
in stressed environments usually expected loss is quite high -- this entails a reduction of unexpected loss.
To some extent, however, this observation might also be a consequence of
choosing a too simple portfolio model and a too simple PIT estimation approach for the purpose of this chapter
as defaults become stochastically independent in our modelling framework when it is used for PIT calculations.
Further investigation is recommended of how to modify ``conditional independence'' modelling in order to
create appropriate levels of correlations for PIT calculations. Promising approaches are the use of latent systematic
factors and the inclusion of contagion effects in the modelling. \citet{Duffie&Eckner&Horel&Saita} and some of
the references therein provide further examples of extending approaches used for PD modelling towards portfolio risk modelling.


\end{document}